\documentclass[12pt,preprint]{aastex}

\singlespace

\newcommand{\Fig}{Figure~}
\newcommand{\Tab}{Table~}
\newcommand{\NH}{N_{\rm H}}

\received{June 11, 2002}
\accepted{August 14, 2002}
\shorttitle{O and Ne interstellar X-ray absorption structures}
\shortauthors{Takei et al.}

\begin{document}

\title{O and Ne K absorption edge structures and interstellar
abundance\\ towards Cyg X-2}
\author{Yoh Takei, Ryuichi Fujimoto and Kazuhisa Mitsuda}
\affil{Institute of Space and Astronautical Science, 3-1-1, Yoshinodai, Sagamihara, Kanagawa, 229-8510, Japan}
\email{takei@astro.isas.ac.jp, fujimoto@astro.isas.ac.jp, mitsuda@astro.isas.ac.jp}
\and
\author{Takashi Onaka}
\affil{Department of Astronomy, School of Science, University of Tokyo, Bunkyo-ku, Tokyo, 113-0033, Japan}
\email{onaka@astron.s.u-tokyo.ac.jp}

\begin{abstract}
We have studied the O and Ne absorption features in the X-ray spectrum
of Cyg X-2 observed with the Chandra LETG.  The O absorption edge is
represented by the sum of three absorption-edge components within the
limit of the energy resolution and the photon counting statistics.
Two of them are due to the atomic O; their energies correspond to two
distinct spin states of photo-ionized O atoms.  The remaining edge
component is considered to represent compound forms of oxide
dust grains. Since Cyg X-2 is about 1.4 kpc above the galactic disk,
the H column densities can be determined by radio (21~cm and
CO emission line) and ${\rm H}_\alpha$ observations with
relatively small uncertainties.  Thus the O abundance relative to H
can be determined from the absorption edges.  We found that the dust
scattering can affect the apparent depth of the edge of the compound
forms. We determined the amplitude of the effect, which we consider is
the largest possible correction factor.  The ratio of column densities
of O in atomic to compound forms and the O total abundance were
respectively determined to be in the range $1.7^{+3.0}_{-0.9}$ to
$2.8^{+5.1}_{-1.5}$ (ratio), and $0.63\pm 0.12$ solar to $0.74 \pm
0.14$ solar (total), taking into account the uncertainties in the
dust-scattering correction and in the ionized H column density. We also
determined the Ne abundance from the absorption edge to be $0.75 \pm
0.20$ solar.  These abundance values are smaller than the widely-used
solar values but consistent with the latest estimates of solar
abundance.
\end{abstract}


\section{Introduction}

The metal abundance in the interstellar medium (ISM) is an important
parameter for the understanding of the chemical evolution of the
Universe.  However, large uncertainties remain in our knowledge of
this abundance.  The solar abundance is sometimes regarded as the
average of our Galaxy.  However, UV absorption lines due to
interstellar gas and optical absorption lines in the atmosphere of
young B-stars have indicated that the metal abundance in the ISM of
our Galaxy is on average about two thirds of the solar abundance
\citep{Savage_Sembach_1996, Snow_Witt_1996}.  However, both methods
contain uncertainties because the former method is only sensitive to
matter in atomic forms, while the abundance of B-star atmospheres may
be different from the ISM.  Indeed, \citet{Sofia_Meyer_2001} recently
claimed that B-star abundances are lower than the ISM abundance and
that the solar abundance is close to the ISM abundance. Moreover, the
O solar abundance itself is not determined well.  Recent values by
\citet{Grevesse_Sauval_1998} and \citet{Holweger_2001} are
significantly lower than the values in the widely used table of
\citet{Anders-Grevesse_1989}.
%In this paper, we quote O and Ne
%abundances assuming the abundance in \citet{Anders-Grevesse_1989} as
%one solar, otherwise noted.

High resolution X-ray spectroscopy of absorption features due to the
interstellar matter is a powerful tool to measure the amount of
interstellar metal.  It is sensitive to both the atomic (gas) and
compound (molecular and dust grain) forms and can distinguish them by the
chemical shifts.  There are absorption lines and edges of many elements
in the X-ray energy range.  We can determine the column densities of
these elements in different forms separately,  integrating towards the X-ray
source.  However, the observation requires both high energy
resolution and good statistics.  \citet{Schatternburg_Canizares_1986}
determined the O column density towards the Crab nebula with the
transmission grating on the Einstein observatory.  They could not
separate atomic and compound forms.  Also they were not able to
determine the O abundance since the H column density
towards the Crab nebula was not known.  \cite{Paerels_etal_2001}
analyzed the O and Ne absorption structures of the
X-ray binary X0614+091 observed with the low energy transmission
grating (LETG) on the Chandra observatory. They successfully separated
the absorption features of O in atomic and compound forms.  They
also determined the abundance ratio of Ne to O.  However, they
could not determine the O abundance because the H
column density toward the source is not known.
\cite{Schulz_etal_2002} determined the column densities of O,
Ne, Mg, Si, and Fe towards Cyg X-1 with the Chandra
high energy transmission grating (HETG) data.  They discussed the
abundances of O, Ne, and Fe.  However, it was necessary to
assume the abundances of all other elements to be solar, in order to
estimate the H column density towards Cyg X-1 from the X-ray
spectrum.

In this paper we analyze the spectrum of Cyg X-2 obtained with the
Chandra LETG and HRC (High Resolution Camera)-I. Cyg X-2 is a Low-Mass
X-ray Binary (LMXB) located at ($l$,$b$)=
($87.33^\circ$,$-11.32^\circ$) and its distance is determined to be
7.2~kpc from the optical counterpart and X-ray burst with mass
ejection \citep{Cowley_Crampton_Hutchings_1979, Smale_1998, Orosz_Kuulkers_1999}.
\cite{Smale_1998} claimed a distance of 11.6 kpc, however, this is
because he assumed the Eddington luminosity for a hydrogen rich
atmosphere.  If we assume a helium rich atmosphere, which is more
realistic for type-I bursts \citep{Lewin_etal_1995}, the distance is
consistent with 7.2~kpc.  Adopting this distance, Cyg X-2 is located
$\sim$ 1.4~kpc above the Galactic disk.  Thus the H column density
towards Cyg X-2 can be estimated from 21 cm (\ion{H}{1}), CO (${\rm
H}_2$), and ${\rm H}_{\alpha}$ (\ion{H}{2}) observations with relatively
small errors.  Because of the featureless X-ray spectrum of the LMXB,
the high brightness, and the well-determined H column density to the
source, Cyg X-2 is one of the best X-ray sources for the study of the
interstellar abundance with X-ray absorption.

In the next section, we will first discuss the H column
density towards Cyg X-2. Then in the following sections we will show
the observational results and discuss the O and Ne abundances and
the uncertainties of the present results.

Throughout this paper, we quote single parameter errors
at the 90~\% confidence level unless otherwise specified.
%we will quote single parameter errors at 90 \% confidence level.

\section{H column density towards Cyg X-2}

Interstellar hydrogen is distributed in three different forms; atomic
(\ion{H}{1}), molecular (H$_2$), and ionic (\ion{H}{2}). 

% atomic hydrogen
The column density of atomic hydrogen is measured by the 21 cm radio
emission. 
In the \ion{H}{1} map of \cite{Dickey_Lockman_1990}, the H column
density ($\NH$) is given at grid points of 0.8 degree separation.
By interpolating between the values of nearby points, the column density
in the direction of Cyg X-2 is estimated to be
$\NH(\mathrm{atom})$=$2.17\times10^{21}~\mathrm{cm^{-2}}$. Since the
scale height of atomic hydrogen is $\sim 140$ pc, the total
column density of our galaxy can be regarded
as the column density to Cyg X-2.  The standard deviation of the $\NH$
values within two degrees of Cyg X-2 is $0.1\times
10^{21}~\mathrm{cm^{-2}}$. The power spectra of the small-scale
spatial fluctuation of the galactic $\NH(\mathrm{atom})$ were
measured for different directions \citep{Crovisier_Dickey_1983,
Dickey_etal_2001}.  On the spatial scales less than 1 deg, the
power-law index of the power spectra is in the range from $-3$ to
$-4$. Assuming the same shape of the power spectrum for the spatial
fluctuations in the vicinity of Cyg X-2, we estimate the rms amplitude
of the spatial fluctuation less than 2 degrees to be $0.12 \times
10^{21}~\mathrm{cm}^{-2}$ and regard this as the 1-$\sigma$ error of
the estimation.

% molecular hydrogen

\cite{Miramond_Naylor_1995} determined the upper limit of molecular hydrogen
column density towards Cyg X-2
from the upper limit of CO emission line in the direction of Cyg X-2
to be $0.05 \times 10^{21}~\mathrm{cm}^{-2}$ (90~\% significance level).
According to them, the contribution of H$_2$ in  $\NH$ 
is smaller than  $0.10 \times 10^{21}~\mathrm{cm}^{-2}$.

% revision starts
% total (HI and H2) column density
The total neutral H column density ($N_\mathrm{H I} + 2N_\mathrm{H_2}$) is
thus estimated to be
$2.17^{+0.22}_{-0.20} \times10^{21}~\mathrm{cm^{-2}}.$
On the other hand, the reddening towards Cyg X-2 is determined to be
$E(B-V) \simeq 0.4$ \citep{McClintock_etal_1984}.
Combining this with the relation,
$N_\mathrm{H I}+2N_\mathrm{H_2}=
5.8\times 10^{21}~E(B-V)~\mathrm{atoms~cm^{-2}~mag^{-1}}$
\citep{Bohlin_etal_1978},
we obtain $N_\mathrm{H I} + 2N_\mathrm{H_2} \simeq
2.3\times 10^{21}~\mathrm{cm^{-2}},$
which is consistent with the above estimation.
% revision ends

% ionized hydrogen
% 
A large fraction \citep[$\sim 97$ \%,][]{Mathis_2000} of ionized H atoms
exists in the warm ionized interstellar medium (WIM) of temperature $T
\sim 8000$ K, distributed in our galaxy with a scale height $h$ of $\sim
900$ pc and a volume filling factor $f$ of  $\gtrsim 0.2$.  In the WIM,
oxygen is effectively ionized through the charge exchange process.  
The number densities of neutral/ionized oxygen and hydrogen 
are related to $N$(\ion{O}{2})$/N$
(\ion{O}{1}) = (8/9) $N$(\ion{H}{2})$/N$(\ion{H}{1})
\citep{Field_Steigman_1971}.  According to more recent model
calculations by \cite{Sembach_etal_2000}, the ionization fraction of
Ne is also similar to that of H; in their `COMPOSITE' model, the
values of $\log (N(X^i)/N(X_{\rm total}))$  where $X=$ H, O, or Ne and $i$
represents the ionization states are, respectively, $-0.73$ and
$-0.09$ for \ion{H}{1} and \ion{H}{2}, $-0.71$ and $-0.09$ for \ion{O}{1} and
\ion{O}{2}, and $-0.70$ and $-0.10$ for \ion{Ne}{1} and \ion{Ne}{2}.
Therefore, $N(X_{\rm total})/N({\rm H}_{\rm total}) =
N(X^0)/N$(\ion{H}{1}) is a good approximation for gas-phase O and Ne and
we do not need to consider ionized H when we calculate atomic O and
Ne abundances.

On the other hand, since interstellar dust grains are thought to
co-exist with the WIM, we need to take into account the column density
of ionized H when we estimate the abundance of O in compound forms.
We estimated it from the H$_\alpha$ emission line intensity in the
direction of Cyg X-2 obtained from the {\it Wisconsin H-Alpha Mapper}
(WHAM) survey \citep{Haffner_etal_prep}
\footnote{http://www.astro.wisc.edu/wham/}.  The intensity is 
$14.6 \pm 3.8$ Rayleighs.
From the line intensity, the total emission measure in the column
towards Cyg X-2 is estimated to be $(1.28 \pm 0.33) \times 10^{20}
(T/8000~\mathrm{K}) ~\mathrm{cm}^{-5}$,
where we applied an 
extinction correction factor of 1.3 assuming 
$R_{\rm v}= 3.1$ \citep{Mathis_1990}. 
Assuming the spatial distribution described above, the emission
measure and the column density are respectively related to 
the local ionized H density in the galactic plane, $n_0$, by,
\[
EM = f \cdot n_0^2 \int_{0}^{\infty}  \left[\exp\left(- \frac{x \sin b}{h}\right) \right]^2 dx
\]
and
\[
\NH(\mathrm{ion}) =  f \cdot n_0 \int_{0}^{d} \exp\left(- \frac{x \sin b}{h}\right)  dx,
\]
where $d$ and $b$ are the distance and the galactic latitude of Cyg
X-2, respectively.  From these we obtain $\NH(\mathrm{ion}) = 0.70
\pm 0.18 \times 10^{21} (f/0.2)^{1/2}~ \mathrm{cm}^{-2}$.
As a nominal value, we adopted
$\NH(\mathrm{ion}) = 0.70\pm0.18\times 10^{21}~ \mathrm{cm}^{-2}$
\cite[$f=0.2,$][]{Mathis_2000} and thus 
$\NH(\mathrm{total}) = 2.87^{+0.25}_{-0.23}\times 10^{21}~ \mathrm{cm}^{-2}.$
Even for $f=1$, $\NH(\mathrm{ion})$ does not exceed 
$1.57 \pm 0.40 \times 10^{21}~ \mathrm{cm}^{-2}.$

%The
%largest uncertainty of $\NH(\mathrm{ion})$ arises from the uncertainty
%of $f$.  Anyway it does not exceed $1.97 \times 10^{21}$ which is
%the value for $f=1$.

In \Tab\ref{tab:Hcolumn}, we summarize the H column densities
estimated in this section.

\section{Analysis and results}

\subsection{Data reduction}

Cyg X-2 was observed with the LETG/HRC-I for 30~ks on 2000 April 24 (obsID
87).  We retrieved the archival data from the CXC (Chandra X-ray Center).
Cyg X-2 is known to show both intensity and spectral
variations.  However, since the X-ray
flux varied only $\sim 20~\%$ during the observation and since we
are interested only in the interstellar absorption features, we
integrated all the data after the standard data screening.

Since the spectrum in the archival data we retrieved is known to 
contain some processing errors\footnote{
see \url{http://asc.harvard.edu/ciao/threads/spectra\_letghrcs/}},
 we reprocessed the data according to the standard methods 
described in {\it threads} in CIAO 2.2.1 (CALDB 2.10).
We first obtained the spectra of the positive and negative orders.
Then we summed them, rebinned
to 0.025~\AA~ bins, and subtracted the background.
The background was only $\sim 2~\%$ of the source counts in the wavelength
region we used in the later analysis. 

\begin{figure}
\epsscale{0.6}
\plotone{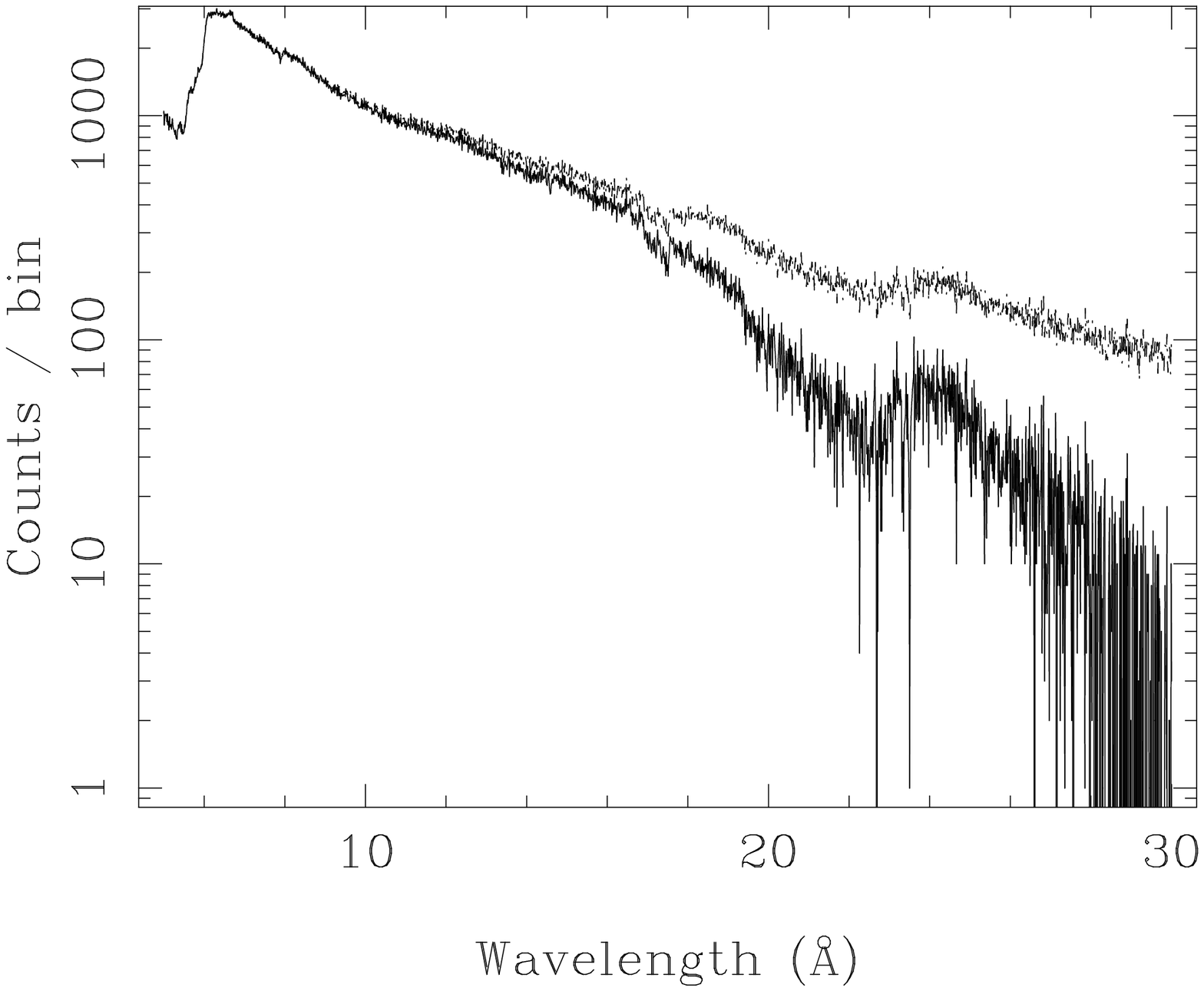}
\caption{The Cyg X-2 spectrum obtained with the Chandra LETG/HRC-I.  The raw
LETG spectrum containing higher order photons is shown with a thin line,
while the spectrum after higher-order subtraction (see text) is shown
with a thick line. 
}
\label{fig_cygx2}
\end{figure}

HRC-I has almost no energy resolution. The spectrum we obtained contains
photons not only of the first order dispersion but also of the higher
orders.  We subtracted the higher order spectra utilizing the method
described in \cite{Paerels_etal_2001}.  The $m$-th order spectrum of
photons with a wavelength of $\lambda$ appears in the wavelength bin,
$\hat{\lambda} = m \lambda$, of the first order spectrum.  Since
the LETG/HRC-I system has no detection efficiency
below the wavelength $\lambda_0$, where $\lambda_0 = 1$~\AA, we can
neglect the contribution of the $m$-th order spectrum in the
wavelength range $\hat{\lambda} < m \lambda_0$.  From the spectrum in
the range $\lambda_0 < \hat{\lambda} < 2\lambda_0$ and the efficiency
ratio of the second to the first order spectra, we can estimate the
second order spectrum in the range $2 \lambda_0 < \hat{\lambda} <
4\lambda_0$ and subtract it from the total spectrum to obtain the
first order spectrum in $ \lambda_0 < \hat{\lambda} < 3\lambda_0$.  In
the same manner, we can subtract higher order spectra from all the
wavelength range.  We calculated the efficiency ratio from the
effective areas of the LETG/HRC-S system provided at the CALDB WWW
page
\footnote{letgs\_NOGAP\_EA\_001031.mod for the 1st order and
letgs\_EA\_001031\_o$m$.mod for $m$-th order, at \url{
http://asc.harvard.edu/cal/Links/Letg/User/Hrc\_QE/EA/high\_orders/ }}
because the higher order efficiencies are not available for the
LETG/HRC-I system and because the efficiency ratios are not
dependent on the detector.  The calibration files we used are
updated compared to those used by \cite{Paerels_etal_2001}.  We show in \Fig
\ref{fig_cygx2} the spectra with and without higher orders.

We estimated the statistical errors of the first order spectrum
obtained with the above method considering the error propagations.
Because of the error propagations, the statistics of the spectral bins
are no longer completely independent. However, the increase of the
statistical errors after the higher order subtraction is smaller than
3\% in the wavelength ranges we use in the following analysis. We thus
treat them to be independent in the spectral fittings.

According to the LETGS calibration report at the CXC
\footnote{see http://cxc.harvard.edu/cal/Links/User/calstatus.html},
the wavelength calibration of LETGS spectra is accurate to 0.02~\%.
This is smaller by an order of magnitude than the statistical errors
in the wavelengths determined in the following analyses.

\subsection{O absorption structure and the optical depth}

\begin{figure}
%\plottwo{f2a.eps}{f2b.eps}
\plotone{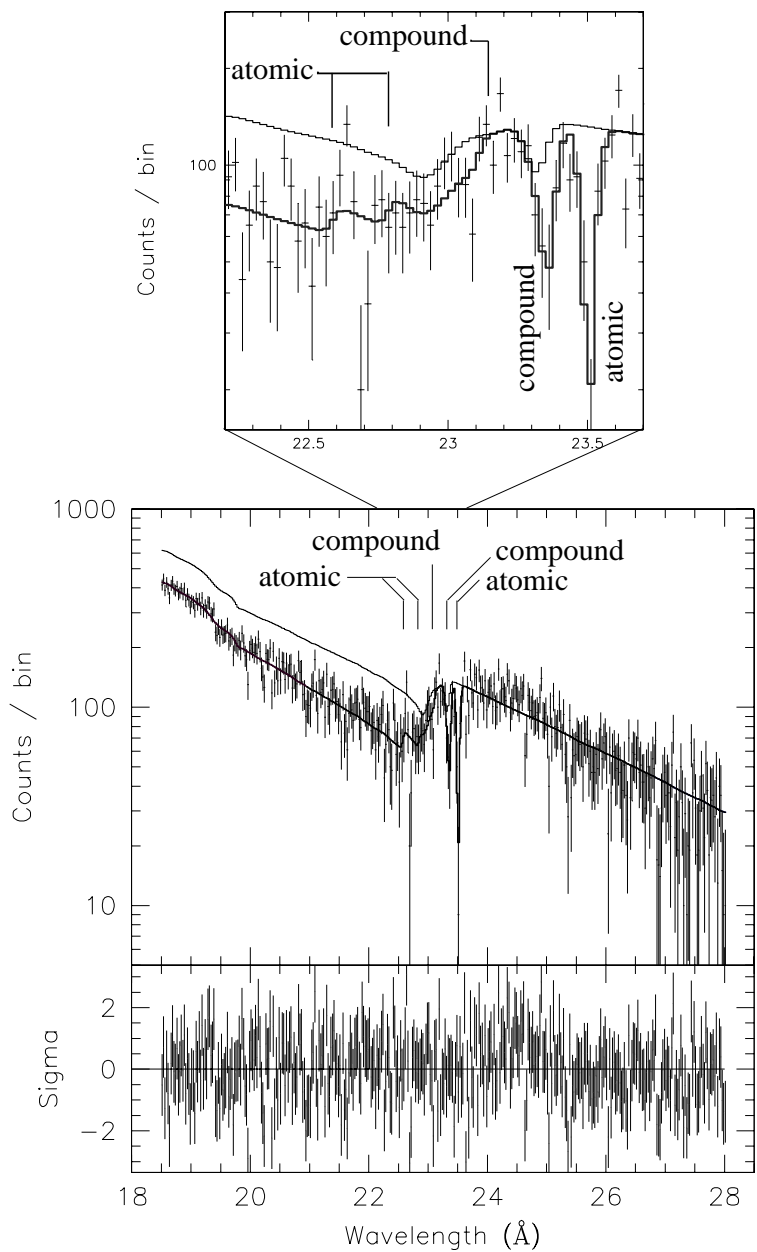}
\caption{
Spectral fit in the wavelength range including the O
absorption structures.  The whole wavelength range used in the
spectral fits is shown in the lower graph, while in the upper graph
the narrow wavelength range containing O edges and absorption lines
is expanded.  In the upper graph and in the upper panel of the lower
graph, the observed 1st order spectrum is shown with data points with
vertical bars which represent 1-$\sigma$ statistical errors. The
best-fit model function of the three edge-component model 
(the last column of Table \ref{tab:fits}) 
convolved with the telescope and the
detector response functions is shown with a thick line. The model
function without O edges and absorption lines is shown with a thin
line, in order to show the instrumental O absorption structure.  Two
absorption lines and an extended O edge clearly exist. In the lower
panel of the lower graph, the residuals of the fit are shown}
\label{fig:Oene}
\end{figure}

In the observed spectrum of Cyg X-2, we clearly found two absorption
lines and an absorption edge near $\lambda\sim$~23~\AA\ besides the
instrumental O absorption line and edge (see \Fig\ref{fig:Oene}).  The
centroid energies of the two absorption lines were determined to be
$23.505 \pm 0.007$~\AA\ and $23.360 \pm 0.014$ \AA, and thus
they were identified with the 1s--2p resonant lines of the atomic and compound O,
respectively.  However, since the absorption lines are saturated
($\tau\gg1$), the absorption column densities cannot be determined
accurately from the absorption lines. We thus focus on the absorption
edge.

The absorption edge structure was extended compared to the instrumental
O K-edge, which suggests that the edge structure consists of
absorption edge components with different edge energies.

A complicated structure is expected for the absorption edge of an
atomic O, firstly because photo-ionized O atoms take
two distinct final states with different total spin numbers, and secondly
because there are many absorption lines which cannot be resolved with
the LETGS near the absorption edges \citep{McLaughlin_Kirby_1998}.
The theoretical values of edge energies calibrated with ground
experiments are 544.03 eV (22.790 \AA, spin 3/2) and 548.85 eV (22.590
\AA, spin 1/2) \citep{McLaughlin_Kirby_1998}.  On the other hand, the
edge energy of the compound O depends on the chemical state.  Thus the
absorption edge of the compound forms can have an extended structure. 
The edge energy must be lower than the edge energy of atomic O,
but higher than the absorption line energy.  According to
\cite{Sevier_1979}, the edge energy of O in metal oxide is
$532.0\pm0.4$ eV ($23.31\pm0.02$ \AA).

All the O edge structures may not be identified because the statistics
and the energy resolution of the present data are limited. Thus, we
first apply a rather simple model to the observed spectrum, then later
try more and more complicated models.

We restricted the wavelength range of model fits
in order to avoid the contributions of absorption edges of other elements.
We did all the analysis described below with several different
wavelength ranges.  We found that  the O absorption optical depth
determined from the analysis is not dependent on the wavelength ranges
of the fits, so long as the lower and upper limits of the range are
within the ranges 17.5--19.0~\AA\ and 27.0--29.0~\AA
, respectively.  Thus we
show here the results with the fitting range of 18.5~\AA--28.0~\AA.

We assumed a power-law function for the continuum spectrum.  We also
included the absorption by neutral matter represented by the {\tt
tbabs} model in the `sherpa' program \citep{Wilms_Allen_McCray_2000} with the O
abundance fixed to 0.  We included an O absorption edge with a
separate absorption edge component represented by {\tt edge} model in
the sherpa program.  We fixed the H column density of the {\tt tbabs}
absorption model to $2.17\times10^{21}~\mathrm{cm^{-2}}$ and fixed the
helium abundance to the solar value.  We performed the fits with the
abundances of other atoms fixed at several different values between
0.5 and 1.5 solar.  We found that the dependence of the best-fit O
absorption optical depths on the abundance of other elements is only
about ten percent of the statistical errors. Thus we only show
the results with the abundance of other elements fixed at the solar
values.

In the first model, we included two oxygen edge components. 
We treated the edge energies and the optical
depths of the two `edge' components as free parameters.  We
represented the two O absorption lines with negative Gaussian
functions. We fixed the centroid wavelengths to 23.505 \AA\  and 
23.360 \AA, and the
widths to 0.04 \AA.  The free parameters of the fits were the two O
edge energies and optical depths, the normalization factor and the
power-law index of the continuum, and the normalization factors of the
two Gaussian absorption lines.

We show the results of the fits in the first data column of \Tab
\ref{tab:fits}.  The best fit energies of the two edges were
$22.76^{+0.13}_{-0.04}$~\AA\ and $23.11^{+0.09}_{-0.03}$~\AA,
respectively.  The former is consistent with the atomic O edge of the
spin 3/2 final state within the statistical error. On the other hand,
the latter is in the middle of the atomic and the instrumental O
absorption edges. Thus, it is likely to represent an O edge of
compound forms.
%The optical depths of the two edges are respectively
%$\tau_\mathrm{atom}=0.41^{+0.10}_{-0.08}$ and
%$\tau_\mathrm{compound}=0.27\pm 0.10$. 
% We estimated the statistical
%error of the sum of the two edge optical depths from the $\chi^2$
%contour map (Fig. \ref{**}) to find $\tau_\mathrm{total} = 0.68\pm
%50.08$.

Next, in order to investigate possible complex structures of the edges
and the model dependence of the absorption optical depths, we
increased the number of edge components one by one. In the fits, we
allowed both the energies and optical depths of all the edges to vary.
In the second and third data columns of \Tab\ref{tab:fits}, the
results are summarized.  The $\chi^2$ value shows a significant
improvement from the double to the three edge models (99.6~\%
confidence with an F-test).  The edge energy of the third edge is
22.58 \AA = 549.1 eV.  This energy is close to 
the edge energy for a final spin 1/2.  In the raw spectrum (\Fig
\ref{fig:Oene}), we can find this edge structure.  On the other hand,
the $\chi^2$ value did not improve from the three-edge to the
four-edge models.  Thus we conclude that the observed O edge
structure can be represented with the three-edge model within the
limit of the present statistics and the energy resolution; two of
the edges represent absorption by atomic O corresponding to the two
different O-ion spin states, and the remaining one represents
absorption by O in compound forms.

Finally, in order to determine the spectral parameters and their
statistical errors for the three-edge model more precisely, we
performed the spectral fit with the three-edge model with the two edge
energies of the atomic O fixed at the theoretical values.  In the last
column of \Tab\ref{tab:fits}, we show the results.  We estimated the statistical
error for the total optical depth for the atomic O edge from
the $\chi^2$ contour map to find the total optical depth of $0.43 \pm 0.14$.  
The optical depth of compound forms is $0.26\pm0.11$.

To estimate the column density of O from the absorption optical depth,
we adopted the absorption cross section of \cite{Verner_Yakovlev_1995}:
$\sigma_\mathrm{abs}=4.98\times10^{-19}~\mathrm{cm^2}$ both for atomic
and compound forms.  We obtained $N_{\rm O, atom} =
8.6\pm2.8\times10^{17}~\mathrm{cm^{-2}}$ and $ N_{\rm O, comp}=
5.2\pm2.2\times10^{17}~\mathrm{cm^{-2}}$.

Assuming the $\NH$
values of $2.17 ^{+0.22}_{-0.20}\times10^{21}~\mathrm{cm^{-2}}$
for atomic O and 
$2.87 ^{+0.25}_{-0.23}\times10^{21}~\mathrm{cm^{-2}}$
for compound forms, 
we estimate the
abundance of O to be $0.47 \pm 0.16 $ solar (atomic), 
$0.21 \pm 0.09$ solar (compound), and $0.68\pm0.13$
solar (total), where we assumed the O solar
abundance of $8.51\times10^{-4}$ \citep{Anders-Grevesse_1989}.
The error of the total abundance was estimated from the $\chi^2$
contour map of the spectral fit, taking into account the difference
of $\NH$ values for gas and compound forms and their systematic errors.

\subsection{Absorption structure of Ne}
%IF IT IS DIFFICULT TO CONVERGE NE EDGE ANALYSIS, WE MAY OMIT NE.

\begin{figure}
\plotone{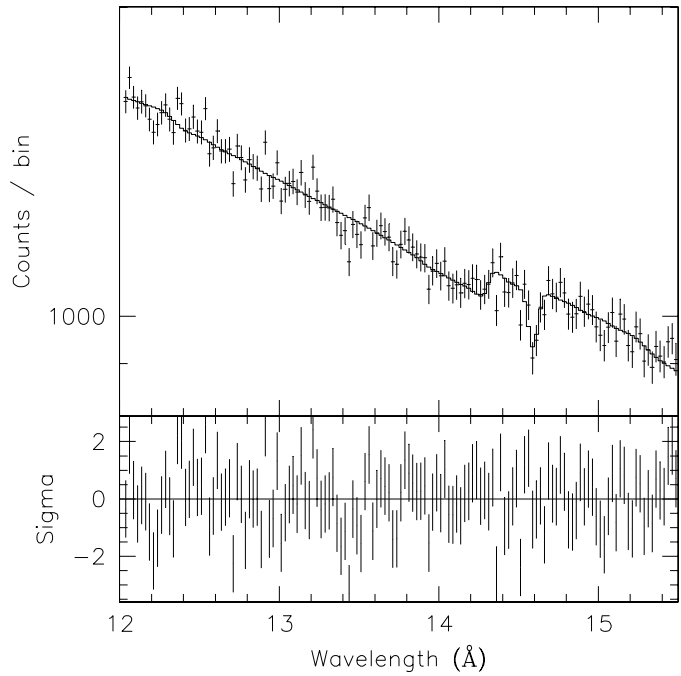}
\caption{Spectral fit in the wavelength range including the Ne absorption
structures. The observed 1st order spectrum with 1-$\sigma$ error bars and
the best fit model function convolved with telescope and detector response
functions are shown in the upper panel, and the residuals of the fit are shown
in the lower panel. 
}
\label{fig:neon}
\end{figure}

In the observed spectrum of Cyg X-2, we also clearly found an
absorption edge at 14.3 \AA, which is identified with the Ne K-edge
(\Fig\ref{fig:neon}). We performed spectral fits to the wavelength
range of 12.0--15.5 \AA.  In addition to the edge, there is an
absorption line at 14.6 \AA.  Because the line wavelength is close to
the edge, we included an absorption line in the model function.
We employed the model function similar to that used for the O
features. We represented the Ne edge with a single edge model.
% and for O.  

In \Tab \ref{tab:neon}, we show the results of the fit. The absorption
edge wavelength was consistent with that of the neutral Ne
\citep[14.25 \AA;][]{Verner_Yakovlev_1995}.  From the optical depth determined from
the fit and the theoretical absorption cross section of \ion{Ne}{1}
\citep{Verner_Yakovlev_1995}, we determined the column density and
abundance of Ne to be $2.00\pm0.51\times10^{17}~\mathrm{cm^{-2}}$
and $0.75 \pm 0.20$ times the solar abundance of
\cite{Anders-Grevesse_1989}.  

Although the absorption-line energy, $14.599 \pm 0.012$ \AA, is close
to the resonant line energy of \ion{Ne}{2}
\citep[14.631 \AA,][]{Behar_Netzer_2002},
it is outside the 90\% statistical error
domain.  Thus we can not unequivocally identify this absorption line.
However, if we assume it is \ion{Ne}{2}, the column
density estimated from the line is $8.0\pm 2.4\times
10^{16}~\mathrm{ cm^{-2}}$, which is about 1/3 of
\ion{Ne}{1}.  Thus if the absorption line is due to \ion{Ne}{2}, \ion{Ne}{2}
can be attributed to the WIM.

\section{Discussion}

We have analyzed the O and Ne absorption structures in the X-ray
spectrum of Cyg X-2 observed with the Chandra LETG/HRC-I.  The O edge
is represented by the sum of three edge components within the limit of
present statistics and energy resolution.  Two edge components
represent absorptions by atomic O; their edge energies are
consistent with the theoretical values corresponding to the two distinct
final O states of different spin numbers.  The edge energy of the
remaining edge is lower than the other two, and is likely to
represent O in compound forms.  From the depth of the edges, we
estimated the O column densities with atomic and compound forms
separately.  

We tried a four-edge component model in the spectral fits.  We find
the sum of the optical depths of all the edge components does not
change significantly from three to four edge models.  Thus we conclude
that the total O column density does not depend significantly on the
number of edge components assumed in the spectral models. We estimated
the atomic, compound, and total O abundances to be $0.47 \pm 0.16$,
$0.21 \pm 0.09$, and $0.68 \pm 0.13$ times the solar abundance of
\cite{Anders-Grevesse_1989} respectively.  The Ne edge is represented
with a single edge model and the Ne abundance is estimated to be
$0.75\pm0.20$ solar.  In the errors of abundance quoted above, we
included the statistical errors of the spectral fits, and the
systematic errors of $\NH$ estimated from the spatial fluctuation of
21cm and ${\rm H}_\alpha$ emissions and the upper limit of CO
emission.

There is an additional systematic error in the abundance of O in compound
forms, arising from the uncertainty in the spatial distribution of the
WIM.  If we consider the extreme case in which its volume filling factor
is 1, the O abundance reduces to $0.16 \pm 0.07$ (compound form) and $0.63\pm0.12$ (total).

%Systematic errors
% =  Uncertainty in NH. 
%    Uncertainty of O edge structure.
%     notice  total O is constant from double to four-edge models.
% effect of dust scattering

% Circumsteller ??
It is very unlikely that the O and Ne absorption edges are associated
with the circumstellar matter in the Cygnus X-2 binary system. Suppose that the
absorption column of $\NH \sim 2\times10^{21} {\rm cm}^{-2}$ is
located in a distance $r$ from the X-ray source. The ionization
parameter, $\xi = L_{\rm X}/(n r^2) \sim L_{\rm X}/(\NH r)$ must
be smaller than $\sim 10$ so that O is neutral
\citep{Kallman_McCray_1982}. With $L_{\rm X} = 10^{38}$ erg sec$^{-1}$, we
find $r\gtrsim 10^{16}$ cm.  This is much larger than the size of the
binary system \citep[$\sim 10^{12}$ cm,][]{Cowley_Crampton_Hutchings_1979}
%{Cowley_etal_1979}).
This is
consistent with the facts that all emission and absorption line
features found in LMXBs are from highly ionized atoms
\citep[e.g.,][]{Asai_etal_2000, Cottam_etal_2001} and that low-ionization
emission lines from the Cyg X-2 system are from near the
companion star \citep{Cowley_Crampton_Hutchings_1979}.
%{Cowley_etal_1979}.

%\subsection{Effect of dust scattering}

About one third of O contributing to the absorption edges is in
compound form.  Dust grains are considered to
contribute to this component.  In such a situation, the optical depth for
dust scattering of X-rays is not negligible compared to the absorption
if we assume a typical dust radius of $\sim 0.1~\mu$m.  Scattered X-rays
form an extended halo on the few arcminute scale
\citep{Hayakawa_1970, Mauche_Gorenstein_1986, Predehl_Schmitt_1995}.
Since the scattering cross section is dependent on the X-ray energy,
the energy spectrum of the unscattered X-rays is modified by
scattering.  However, if multiple scattering is negligible and if the spatial
distribution of dust is uniform over the spatial scale corresponding
to the dust halo, the energy spectrum of total photons, i.e., the sum of the
unscattered core and the scattered halo, is not modified by dust
scattering.  This is not the case for the present energy
spectrum obtained with the LETG, since we derived the energy spectrum only from
the central $\sim$ 1 arc second of the X-ray image.  Because
the scattering cross sections show anomalous features around absorption
edges \citep{Mitsuda_etal_1990}, the dust scattering will affect the
estimation of the absorption column density; the scattering cross
section is {\it smaller} at the energy just above the edge (at the
wavelength shorter than the edge), while the absorption cross section
is {\it larger}.  This leads to an underestimation of the O absorption
column density.  We estimated the amplitude of this effect in the
following way.

The total scattering cross section of an X-ray photon of a wavelength
$\lambda$ with a dust grain of a radius $a$ is given by
\[
\sigma_\mathrm{dust,sc}=2\pi a^2 \left(\frac{2\pi a}{\lambda}\right)^2
\left|\frac{N r_0 \lambda^2}{2\pi} (f_1+{\rm i}f_2)\right|^2,
\]
under the Rayleigh-Gans approximation \citep{van_de_Hulst_1957}, where $N$ is
the number of atoms per unit volume, $r_0$ is the classical electron
radius, and $f_1$ and $f_2$ are the atomic scattering factors 
\citep{Henke_Gullikson_Davis_1993}.  
The Rayleigh-Gans approximation is valid for
$(2\pi a/\lambda) | N r_0 \lambda^2/(2\pi)
(f_1+{\rm i}f_2) | \ll 1$.

%Notice that this is cross section {\it per a dust}, not
%{\it per an atom}.  
Because the factor $|f_1 + {\rm i}f_2|$ becomes small
near the absorption edge, the dust scattering cross section is reduced
at the edge.  Assuming typical
chemical compositions and densities of dust grains
($\mathrm{FeSiO_4}$, $\mathrm{FeSiO_3}$, $\mathrm{Mg_2SiO_4}$
and $\mathrm{MgSiO_3}$), we estimate the reduction of the 
scattering cross section at the O edge to be in the range,
\[
\Delta \sigma_\mathrm{dust,sc}=(0.53-0.93)\times10^{-10}
\left(\frac{a}{0.1~\mathrm{\mu m}}\right)^{4}~\mathrm{cm}^2.
\]
Assuming the number density of dust to be proportional to $\propto
a^{-3.5}$ with a cut off at $a_\mathrm{max}$ \citep{Mathis_Rumpl_Nordsieck_1977},
we calculate the reduction of the scattering cross section per O
atom,
\[
\Delta \sigma_\mathrm{sc}=(0.74-1.44)\times10^{-19}
\left(\frac{a_\mathrm{max}}{0.1~\mathrm{\mu m}}\right) ~
\mathrm{cm}^2.
\] Thus the reduction of scattering cross section can be
a few tens of percent of the absorption cross section,
$\sigma_\mathrm{abs}=4.98\times10^{-19}~\mathrm{cm^2}$.  The maximum
grain radius contributing to X-ray scatterings has been estimated from
the spatial size of dust scattering halos.
\citet{Mauche_Gorenstein_1986} and \citet{Mitsuda_etal_1990} suggested
$a_\mathrm{max} = 0.05 - 0.1~\mathrm{\mu m}$, but recently
\cite{Witt_Smith_Dwek_2001} suggested the presence of grains with radii as large
as $0.5\mathrm{\mu m}$.  Adopting $a_\mathrm{max} = 0.1~\mathrm{\mu
m}$, and assuming that O of compound forms are all in dust grains, the
column density of O of compound forms increases from
$5.2\times10^{17}~\mathrm{cm^{-2}}$ to $(6.1 - 7.3) \times
10^{17}~\mathrm{cm^{-2}}$.  At the O edge energy, the Rayleigh-Gans
approximation is valid only for $a_\mathrm{max}~\lesssim
0.1~\mathrm{\mu m}$ and the scattering cross section saturates
at $a_\mathrm{max}\gtrsim 0.1~\mathrm{\mu m}$
\citep{Alcock_Hatchett_1978,Smith_Dwek_1998}.  Thus the correction
factor does not increase for larger grains and the correction factor
we applied above can be regarded as the maximum correction.

The O abundance values calculated for four different cases 
(with and without correction for dust scattering, and two different
values of the volume filling factor of the WIM) are summarized
in \Tab\ref{tab:summary}.  The total O abundance is
between $0.63\pm 0.12$ and $0.74 \pm 0.14$ times solar.  
On the other hand the Ne
abundance is $0.75 \pm 0.20$.  Our results are more consistent with
the recent solar abundances than the most widely used `old' values.
For example , the O and Ne solar abundances by \cite{Holweger_2001}
are 0.65 and 0.81 times the values in \cite{Anders-Grevesse_1989},
respectively.

Another important parameter we can estimate from the present X-ray
observation is the gas to dust ratio of the O column densities, which
is estimated to be $2.2^{+3.9}_{-1.1}$, $2.8^{+5.1}_{-1.5}$,
$1.7^{+3.0}_{-0.9}$, and $2.2^{+4.0}_{-1.1}$ for cases 1 to 4 of
\Tab\ref{tab:summary}, respectively.  This should be compared with the
value estimated from the gas abundance from interstellar UV absorption
lines and the assumption that the total abundance is solar.  Adopting
the O gas abundance of \cite{Cartledge_Meyer_Lauroesch_2001} and solar abundances
by \cite{Anders-Grevesse_1989} and \cite{Holweger_2001}, we obtain 0.7
and 1.7, respectively. Our result is again consistent with the latest
solar abundance value.
 
For O, the present X-ray result contains an additional uncertainty due
to the correction for dust scattering.  We can avoid this if we
include the spectrum from the scattering halo in the analysis.  This
requires non-dispersive high resolution spectrometers such as
microcalorimeters, which will be used on board future X-ray missions,
such as Astro-E2.

\acknowledgements We thank J. Ichimura and K. Masai for discussions on
the O and Ne absorption structure, P. Edwards for his careful
review of the manuscript.
We appreciate the H$_\alpha$ survey of the Wisconsin H-Alpha Mapper funded by the
National Science Foundation.
We are also grateful to the anonymous referee
for his valuable comments, in particular, suggesting
the importance of \ion{H}{2} in the H column density.

%\section{Discussion}

%\section{Conclusion}

\begin{deluxetable}{lll}
%\rotate
\tablecolumns{3}
%\tablewidth{0pt}
\tablecaption{Hydrogen column density to Cyg X-2\label{tab:Hcolumn}
}
%\tabletypesize{\scriptsize}
\tablehead{
\colhead{Form of H} & \colhead{} & 
\colhead{H column density}
}

\startdata 
\ion{H}{1} &        & $2.17 \pm 0.20 \times10^{21}~\mathrm{cm^{-2}}$ \\
${\rm H}_{2}$&      & $< 0.10 \times 10^{21}~\mathrm{cm}^{-2}$\\
\ion{H}{2} &(nominal)$^{(3)}$ & $0.70\pm 0.18 \times10^{21}~\mathrm{cm^{-2}}$ \\
\ion{H}{2} &(maximum)$^{(4)}$ & $1.57 \pm 0.40 \times10^{21}~\mathrm{cm^{-2}}$ \\
\\
\ion{H}{1}+${\rm H}_{2}$~$^{(1)}$ & & $2.17^{+0.22}_{-0.20} \times10^{21}~\mathrm{cm^{-2}}$ \\
All forms $^{(2)}$ &(nominal)$^{(3)}$ & $2.87^{+0.25}_{-0.23} \times10^{21}~\mathrm{cm^{-2}}$ \\
All forms $^{(2)}$ &(maximum)$^{(4)}$ & $3.74 \pm 0.45\times10^{21}~\mathrm{cm^{-2}}$ \\

\enddata
\\
\tablenotetext{(1)}{Column density to be used for the calculation of O in the atomic form and Ne.}
\tablenotetext{(2)}{Column density to be used for the calculation of O in the compound forms.}
\tablenotetext{(3)}{Volume filling factor of 0.2 is assumed for WIM.}
\tablenotetext{(4)}{Volume filling factor of 1 is assumed for WIM as an extreme case.}
\end{deluxetable}

\begin{deluxetable}{llcccc}
\rotate
\tablecolumns{6}
\tablewidth{0pt}
\tablecaption{Results of spectral fits of O K-edge absorption features\label{tab:fits}
}
\tabletypesize{\scriptsize}
\tablehead{
\colhead{Parameters} & \colhead{} &
\colhead{Two Edges} &
\colhead{Three Edges} & \colhead{Four Edges$^{(1)}$} &
\colhead{Three with two fixed$^{(2)}$}
}

\startdata

\underline{Power-law}\\
% normalization &(photons s$^{-1}$ keV$^{-1}$ cm$^{-2}$@1keV)
%     & $1.63^{+0.17}_{-0.03}$ & $1.68\pm0.04$ & 1.67 & $1.66\pm0.05$   \\
 normalization &(photons s$^{-1}$ keV$^{-1}$ cm$^{-2}$@555.2 eV)
     & $2.52\pm0.07$ & $2.53\pm0.07$ & 2.53 & $2.53\pm0.07$   \\
 photon index &  
     & $0.74\pm0.17$ & $0.69^{+0.18}_{-0.16}$& 0.70 & $0.71\pm0.17$  \\
\\
\underline{Absorption line 1}\\
 Centroid &(eV) &  527.48 &  527.48 &  527.48 &  527.48\\ 
          &(\AA) &  (23.51)&  (23.51)&  (23.51)&  (23.51)\\
 Equivalent width& (eV)& $1.41 \pm 0.24$& $1.41\pm0.24$& 1.41 & $1.41\pm0.24$ \\
\\
\underline{Absorption line 2}\\
 Centroid &(eV) &  530.76 &  530.76 &  530.76 &  530.76 \\
          &(\AA)&  (23.36)&  (23.36)&  (23.36)&  (23.36)\\
 Equivalent width & (eV)&$0.97 \pm 0.28$& $0.97\pm0.28$& 1.0 & $0.97\pm0.28$ \\
\\
\underline{Absorption Edge 1}\\
 Edge energy &(eV) & $536.4^{+0.8}_{-1.9}$ & 536.2$^{(3)}$
                     & 536.2 & $536.1\pm1.5$\\
             &(\AA)& ($23.11^{+0.09}_{-0.03})$  & (23.12)
                     & (23.13) & ($23.13\pm0.06$)\\
 Optical depth &   & $0.27\pm0.10$ & $0.24\pm0.12$ 
             & 0.24 & $0.26\pm0.11$  \\
\\
\underline{Absorption Edge 2}\\ 
 Edge energy &(eV) & $544.8^{+0.8}_{-3.1}$ & 542.3$^{(3)}$
             & 542.3  &  544.03\\
    &(\AA)& ($22.76^{+0.13}_{-0.04}$)  & (22.86)& (22.86)& (22.79)\\
 Optical depth &   & $0.41^{+0.10}_{-0.08}$ &$0.24^{+0.29}_{-0.13}$ & 0.19 & $0.23^{+0.16}_{-0.11}$\\
\\
\underline{Absorption Edge 3}\\
 Edge energy &(eV) &  & 549.1$^{(3)}$ & 544.9 & 548.85\\
     &(\AA)&  & (22.58)& (22.75) & (22.59)\\
 Optical depth &   &  & $0.22^{+0.13}_{-0.15}$ & 0.09 & $0.20^{+0.15}_{-0.17}$\\
\\
\underline{Absorption Edge 4}\\
 Edge energy &(eV) &  &  & 549.2      \\
       &(\AA)&  &  & (22.58)    \\
 Optical depth&    &  &  & 0.18     \\
\\
$\chi^2$/dof &     &400.45/371 & 397.40/369 & 397.163/367 & 398.21/371     \\
\enddata
\\
\tablenotetext{(1)}{Statistical errors are not estimated for this model.}
\tablenotetext{(2)}{Edge energies of atomic O are fixed at the theoretical values.}
\tablenotetext{(3)}{Statistical errors are not estimated 
because the edge energies are strongly coupled.}
\end{deluxetable}

\begin{deluxetable}{llc}
%\rotate
\tablecolumns{3}
\tablewidth{0pt}
\tablecaption{Results of spectral fits of Ne K-Edge\label{tab:neon}}	
\tabletypesize{\scriptsize}
\tablehead{
\colhead{Parameters} & \colhead{} &
\colhead{Best fit value}
}

\startdata

\underline{Power-law}\\
 normalization &(photons s$^{-1}$ keV$^{-1}$ cm$^{-2}$@1keV)
     & $1.685\pm0.023$ \\
 photon index &  
     & $1.328^{+0.070}_{-0.072}$ \\
\\
\underline{Absorption line} \\
 Centroid &(eV) &  $849.24\pm0.68$ \\
          &(\AA) &  ($14.559\pm0.012$) \\
 Equivalent width& (eV)& $0.541\pm0.160$ \\
\\
\underline{Absorption Edge} \\
 Edge energy &(eV) & $866\pm5$ \\
             &(\AA)& ($14.31^{+0.08}_{-0.09}$) \\
 Optical depth &   & $0.059\pm0.015$ \\
\\
$\chi^2$/dof &     & 149.724/133
\enddata
\end{deluxetable}

\begin{deluxetable}{llcccc}
%\rotate
\tablecolumns{3}
%\tablewidth{0pt}
\tablecaption{Summary of interstellar O and Ne abundances
\label{tab:summary}
}
%\tabletypesize{\scriptsize}
\tablehead{
\colhead{Atom} & \colhead{Form} & 
\colhead{Case 1$^{\rm a}$} & \colhead{Case 2$^{\rm b}$} & 
\colhead{Case 3$^{\rm c}$} & \colhead{Case 4$^{\rm d}$} 
}

\startdata 

O & gas phase & $0.47 \pm 0.16 $ & $\rightarrow$ 
              & $\rightarrow$& $\rightarrow$ \\
  & compound  & $0.21 \pm 0.09 $ & $0.16 \pm 0.07 $ 
              & $0.27 \pm 0.12 $ & $0.21 \pm 0.09 $ \\
  & total     & $0.66 \pm 0.12 $ & $0.63 \pm 0.12 $ 
              & $0.74 \pm 0.14 $ & $0.68 \pm 0.14 $ \\
Ne&           & $0.75 \pm 0.20 $ & & & \\
\enddata
\tablecomments{All abundance values are in units of the solar values of
\cite{Anders-Grevesse_1989}.}
\tablenotetext{a}{Case 1: Volume filling factor of 0.2 is assumed for the WIM. No correction is made for dust scattering.}
\tablenotetext{b}{Case 2: Volume filling factor of 1.0 is assumed for the WIM as an extreme case. No correction is made for dust scattering.}
\tablenotetext{c}{Case 3: Volume filling factor of 0.2 is assumed for the WIM. 
Correction which is considered to be the maximum is made for dust scattering.}
\tablenotetext{d}{Case 4: Volume filling factor of 1.0 is assumed for the WIM as the extreme case. Correction which is considered to be the maximum is made for dust scattering.}
\end{deluxetable}

\end{document}